\documentclass[a4paper,11pt]{article}
\usepackage{jcappub} 

\usepackage[T1]{fontenc} 

\usepackage{graphicx, epsfig, amssymb} 
\usepackage{amsmath, amsfonts}
\usepackage{bm} 

\def\nn{\nonumber}
\def\be{\begin{equation}}
\def\ee{\end{equation}}
\def\beq{\begin{eqnarray}}
\def\eeq{\end{eqnarray}}

\def\mf{\mathfrak}
\newcommand{\PBH}{{\mbox{\tiny PBH}}}
\newcommand{\DM}{{\mbox{\tiny DM}}}

\title{\boldmath Tidal capture of a primordial black hole by a neutron star: implications for constraints on dark matter}

\author[a,b]{Paolo Pani}
\author[b]{Abraham Loeb}

\affiliation[a]{CENTRA, Departamento de F\'{\i}sica, 
Instituto Superior T\'ecnico, Universidade de Lisboa,
Avenida Rovisco Pais 1, 1049 Lisboa, Portugal.}
\affiliation[b]{Institute for Theory and Computation, Harvard-Smithsonian
CfA, 60 Garden Street, Cambridge MA 02138, USA}

\emailAdd{paolo.pani@tecnico.ulisboa.pt}
\emailAdd{aloeb@cfa.harvard.edu}

\abstract{
In a close encounter with a neutron star, a primordial black hole can get gravitationally captured by depositing a considerable amount of energy into nonradial stellar modes of very high angular number $l$. If the neutron-star equation of state is sufficiently stiff, we show that the total energy loss in the point-particle approximation is formally divergent. Various mechanisms --~including viscosity, finite-size effects and the elasticity of the crust~-- can damp high-$l$ modes and regularize the total energy loss. Within a short time, the black hole is trapped inside the star and disrupts it by rapid accretion. Estimating these effects, we predict that the existence of old neutron stars in regions where the dark-matter density $\rho_\DM\gtrsim 10^2 ({\sigma}/{\rm km~s^{-1}}){\rm GeV~cm^{-3}}$ (where $\sigma$ is the dark-matter velocity dispersion)
limits the abundance of primordial black holes in the mass range $10^{17}{\rm g}\lesssim m_\PBH\lesssim 10^{24}{\rm g}$, which was previously unconstrained. In combination with existing limits, our results suggest that primordial black holes cannot be the dominant dark matter constituent. 
}

\begin{document}
\maketitle
\flushbottom
\section{Introduction}
Primordial black holes (PBHs) might arguably be the most natural candidates to solve the dark matter (DM) puzzle. They are cold, weakly-interacting, and do not require extensions of the Standard Model of particle physics. Since being proposed by Hawking in 1971~\cite{Hawking:1971ei} (see also \cite{1969ApANDSS...4..301Z,Carr:1974nx}), substantial effort has been put into detecting these hypothetical objects or ruling them out. 

PBHs could have formed across a wide range of masses as a result of the evolving density and horizon scale of the early Universe~\cite{Carr:2009jm,Carr:2005bd}.
Light PBHs with a mass $m_\PBH<10^{15}{\rm g}$ should have already Hawking-evaporated by the present epoch~\cite{Hawking:1974sw} and observations of the extragalactic photon background put very stringent constraints on the DM fraction in PBHs with $m_\PBH\lesssim10^{17}{\rm g}$~\cite{Carr:2009jm}. On the other hand, strong constraints are in place for nonevaporating PBHs with a mass $m_\PBH \gtrsim 4\times10^{24}{\rm g}$, based on a variety of dynamical~\cite{Quinn:2009zg,0004-637X-516-1-195}, microlensing~\cite{MACHO,EROS,GRB,QSO,Griest:2013aaa} and other astrophysical~\cite{Ricotti:2007au,Mack:2006gz,Kesden:2011ij,Pani:2013hpa} effects (see Ref.~\cite{Carr:2009jm} for an overview). 

Despite extensive theoretical and experimental effort, the range $10^{17}{\rm g}\lesssim m_\PBH\lesssim 10^{24} {\rm g}$ is still poorly constrained\footnote{Recent constraints include those based on femtolensing~\cite{Barnacka:2012bm} and on PBHs in globular clusters~\cite{Capela:2013yf,Capela:2012jz}. However, the former ignored finite-source effects and are currently being revised~\cite{Barnacka}, whereas the latter are weakened by the lack of evidence for DM in globular clusters~\cite{Conroy:2010bs}.}. In this mass range, light PBHs could in principle explain the DM without a need for exotic, weakly-interacting particles. 

Here we consider a novel phenomenon that can be used to derive stringent theoretical constraints on the range $10^{17}{\rm g}\lesssim m_\PBH\lesssim 10^{24} {\rm g}$, the only one that remains phenomenologically allowed to date. We show that in a close encounter with a neutron star (NS), light PBHs can be tidally captured by depositing a surprisingly large amount of energy into the nonradial stellar modes. 
The energy loss by tidal heating is enhanced only when the impact parameter is sufficiently small so that the PBH travels very close or within regions in which the matter equation of state (EOS) is sufficiently stiff. This new phenomenon is associated to the excitation of nonradial stellar modes with high angular number $l$.  
In the absence of a mechanism that can quench such modes efficiently, a PBH would be captured within the NS core in a time scale much shorter than the NS lifetime, and would eventually disrupt the star by rapid accretion~\cite{Giddings:2008gr,Capela:2013yf}. Thus, the mere observation of NSs in DM-rich environments poses a theoretical limit on the density of PBHs. 

The rest of this paper is divided as follows. In Section~\ref{sec:tidal} we discuss the energy deposited in nonradial stellar modes during a close encounter between a NS and a small compact object that interacts purely gravitationally and does not get tidally disrupted. Under certain conditions we find that that the total energy loss is formally divergent and in Section~\ref{sec:cutoff} we discuss various mechanisms that can be advocated to regularize the result. Section~\ref{sec:capture} is devoted to review the tidal capture rate of a point particle by a NS. These results are then used in Section~\ref{sec:limits} to derive stringent limits on the DM fraction in PBHs. We conclude in Section~\ref{sec:conclusions} with some discussion and possible extensions of our work. Some details of the computation are presented in Appendix~\ref{app:normalmodes}.

\section{Tidal heating in a PBH-NS close encounter}\label{sec:tidal}
\subsection{Energy loss by tidal heating}
We consider the encounter of a PBH\footnote{The results of this section apply to any small compact object (e.g. a MACHO) as long as the latter does not get tidally disrupted during the close encounter with the NS. Nonetheless, since our ultimate goal is to discuss the interaction of a NS with a PBH, we shall simply refer to the small compact object as PBH.} with mass $m_\PBH$ and a NS with mass $M\gg m_\PBH$ and radius $R$. For simplicity, we focus on head-on collisions, although our results can be easily generalized to arbitrary orbits with similar results. During a close encounter, the PBH will deposit energy in nonradial acoustic modes of the star. Press \& Teukolsky studied an analogous process for a NS-NS encounter~\cite{1977ApJ...213..183P} (see also Ref.~\cite{1986ApJ...310..176L}) and found that the total energy loss reads (using $G=c=1$ units)
\begin{equation}
 \Delta E=\frac{m_\PBH^2}{R}\sum_{l=2}^\infty\left(\frac{R}{R_{\rm min}}\right)^{2l+2} T_l\,,\label{PT}
\end{equation}
where $R_{\rm min}$ is the periastron distance of a parabolic orbit, $l$ is the multipolar index of the stellar modes and $T_l$ is a function of the masses of the objects, the radius $R$ and the periastron distance~\cite{1977ApJ...213..183P} (see also Refs.~\cite{1994ApJ...426..688R,Lai:1993di} for extensions). 

Typically, the minimum separation in a NS-NS encounter is dictated by the Roche radius within which the smallest object is tidally disrupted~\cite{Shapiro:1983du}, implying $R_{\rm min}\gtrsim 3 R$. Since $T_l$ are roughly the same for any $l$~\cite{1977ApJ...213..183P}, the sum in Eq.~\eqref{PT} converges quickly, and $\Delta E\sim m_\PBH^2/R$. However, a PBH does not get tidally disrupted; it can reach the NS surface and even travel \emph{through} the star. One might therefore expect that the energy exchange be much larger, because higher multipoles are relevant. 

To examine the role of high multipoles, we adopt an approach which is commonly used in seismology~\cite{seismology}. Considering a spherically symmetric, perfect-fluid star with barotropic pressure $P=P(\rho)$, and a point mass $m_\PBH$ travelling radially, we compute the seismic energy deposited into the stellar normal modes~\cite{1989nos..book.....U,Kokkotas:1999bd}. The point-particle approximation is particularly well suited here, because in the range of interest $m_\PBH/M\lesssim 10^{-8}$. Our treatment is therefore Newtonian.
When considering the motion in the interior, we neglect other effects such as accretion and dynamical friction~\cite{Capela:2013yf} because, as we show, they are subdominant relative to tidal heating.

It is convenient to perform a normal-mode decomposition of the displacement of a fluid element~\cite{seismology},
\begin{equation}
 \boldsymbol{s}(\boldsymbol{x},t)=-{\rm Re}\sum_{{\mathfrak n}l}\frac{c_{{\mathfrak n}l}}{\omega_{{\mathfrak n}l}^2} e^{i\omega_{{\mathfrak n}l}t}\boldsymbol{s}_{{\mathfrak n}l}\,, \label{displacement}
\end{equation}
where $\boldsymbol{s}_{{\mathfrak n}l}$ denotes the eigenfunction with overtone ${\mathfrak n}$ and harmonic index $l$ and whose eigenfrequency is $\omega_{{\mathfrak n}l}$. The normal-mode equations are given in Appendix~\ref{app:normalmodes} (see also Refs.~\cite{seismology,1989nos..book.....U,Kokkotas:1999bd,GRL:GRL7294}). Here $c_{{\mathfrak n}l}$ is the excitation coefficient of the $({\mf n},l)$ mode, obtained through a convolution of the source force density $\boldsymbol{f}(\boldsymbol{x},t)$ with the eigenfunctions:
\begin{equation}
  c_{{\mathfrak n}l}=\int dt \,e^{-i\omega_{{\mathfrak n}l}t}\int_{\rm star} d\boldsymbol{x}^3\partial_t \boldsymbol{f}(\boldsymbol{x},t)\cdot\boldsymbol{s}_{{\mathfrak n}l}^*(\boldsymbol{x})\,, \label{ck}
\end{equation}
where the time integration is performed in the interval when the source is active.
The modified seismic energy is the sum over the excitation coefficients~\cite{seismology}
\begin{equation}
 \Delta E\equiv\sum_{{\mathfrak n}l}{E}_{{\mathfrak n}l}=\frac{1}{2}\sum_{{\mathfrak n}l}\frac{|c_{{\mathfrak n}l}|^2}{\omega_{{\mathfrak n}l}^2}\,. \label{seismic}
\end{equation}
Once the source is specified, the knowledge of the normal modes allows us to compute $c_{{\mathfrak n}l}$ and the modified seismic energy through Eq.~\eqref{seismic}. Note that $\Delta E$ is a conservative lower bound on the total deposited energy~\cite{seismology}.

We adopt the point-particle approximation~\cite{Luo:2012pp}
\begin{equation}
 \boldsymbol{f}(\boldsymbol{x},t)\equiv -\rho(\boldsymbol{x})\boldsymbol{\nabla}\Phi=  m_\PBH\rho(\boldsymbol{x})\boldsymbol{\nabla}\frac{1}{|\boldsymbol{x}-\boldsymbol{x}_p(t)|}\,,\label{fsource}
\end{equation}
where $\Phi$ is the gravitational potential, $\boldsymbol{\nabla}=\boldsymbol{\hat{r}}\partial_r+\boldsymbol{\nabla_1}$, $\boldsymbol{\nabla_1}=\boldsymbol{\hat{\theta}}\partial_\theta+\boldsymbol{\hat{\phi}}(\sin\theta)^{-1}\partial_\phi$ and $\boldsymbol{x}_p(t)=(r_p(t),0,0)$ is the location of the PBH. The source term can be expanded in Legendre polynomials as
\begin{equation}
 \boldsymbol{f}(\boldsymbol{x},t)= m_\PBH\rho(\boldsymbol{x})\boldsymbol{\nabla}\left\{\begin{array}{c}
                                                        \sum_l \frac{r^l}{r_p^{l+1}} P_l(\cos\theta)\qquad r_p(t)>r \\
                                                        \sum_l \frac{r^l_p}{r^{l+1}} P_l(\cos\theta)\qquad r_p(t)<r
                                                       \end{array}
\right.
\end{equation}
where we have used the fact that $\theta_p=\phi_p=0$ after a suitable rotation of the axis.
This source excites only spheroidal modes~\cite{seismology,1989nos..book.....U}, which can be conveniently decomposed in a basis of spherical harmonics,
\begin{equation}
 \boldsymbol{s}_{{\mathfrak n}l}(\boldsymbol{x})=U_{{\mathfrak n}l}(r)Y_{lm}(\theta,\phi)\boldsymbol{\hat{r}}+\frac{V_{{\mathfrak n}l}(r)}{\sqrt{l(l+1)}}\boldsymbol{\nabla_1} Y_{lm}(\theta,\phi)\,,
\end{equation}
with no summation over $l$. Due to the symmetry of the problem, only the $m=0$ modes are relevant and there is no explicit dependence on the coordinate $\phi$.

Inserting the expansion above into Eq.~\eqref{ck}, using Eq.~\eqref{fsource} and the orthonormality properties of the spherical harmonics, we can compute $c_{{\mathfrak n}l}$. For the external motion we obtain
\begin{eqnarray}
 c_{{\mathfrak n}l}^{\rm ext}&=&-m_\PBH \int_{-\infty}^{t_R}dt \frac{l(l+1)v_p(t)}{r_p^{l+2}(t)}e^{-i\omega_{{\mathfrak n}l}t}\int_0^R dr\rho(r) r^{l+1}\left(U_{{\mathfrak n}l}+\frac{\sqrt{l(l+1)}}{l} V_{{\mathfrak n}l}\right)\,, \label{cext}
\end{eqnarray}
where $r_p(t<t_R)=\left(R^{3/2}+3 \sqrt{M/2} (t_R-t)\right)^{2/3}$, $v_p=\partial_t{r_p}$, and $t_R$ marks the time at which the particle reaches the surface of the star. For simplicity, to compute the excitation coefficients we have assumed that the PBH is at rest at infinity. This is a conservative assumptions because the deposited energy increases if the initial kinetic energy of the PBH is nonzero. This assumption will be relaxed in Section~\ref{sec:capture} when computing the capture rate, which instead decreases for large initial velocities.

The response when the PBH travels within the NS is more involved and it has to be divided in the case $r_p(t)<r$ and $r_p(t)>r$. The final result reads
\begin{eqnarray}
 c_{{\mathfrak n}l}^{\rm int}&=&-m_\PBH \int_{t_R}^{t_0}dt \frac{l(l+1)v_p(t)}{r_p^{l+2}(t)}e^{-i\omega_{{\mathfrak n}l}t}\int_0^{r_p(t)} dr\rho(r) r^{l+1}\left(U_{{\mathfrak n}l}+\frac{\sqrt{l(l+1)}}{l} V_{{\mathfrak n}l}\right)\nn\\
 &&-m_\PBH \int_{t_R}^{t_0}dt\, l(l+1) v_p(t) r_p^{l-1}(t) e^{-i\omega_{{\mathfrak n}l}t}\int_{r_p(t)}^Rdr\frac{\rho(r)}{r^l}\left(U_{{\mathfrak n}l}-\frac{\sqrt{l(l+1)}}{l+1} V_{{\mathfrak n}l}\right)\,, \label{cint}
\end{eqnarray}
where $t_0>t_R$ is the time the particle reaches the center of the star and $r_p(t>t_R)$ is obtained solving the equation of motion of a point particle travelling through a density distribution $\rho(r)$.
Finally, the seismic energy is computed through Eq.~\eqref{seismic} with $c_{{\mathfrak n}l}=c_{{\mathfrak n}l}^{\rm int}+c_{{\mathfrak n}l}^{\rm ext}$, and multiplying the result by a factor $2$ to account for the symmetric motion from the center to the exterior of the NS.

%

 \begin{figure}[ht]
 \begin{center}
\epsfig{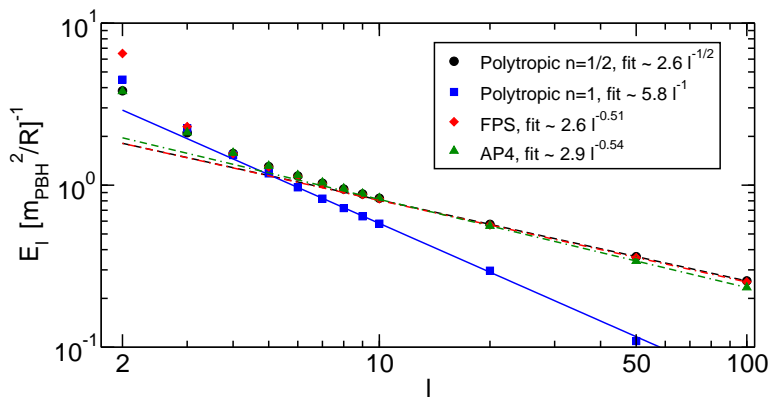}
\caption{\label{fig:El}
Multipolar contributions ${E}_{{\mathfrak n}l}$ to the total energy $\Delta E=\sum_{{\mf n}l}{E}_{{\mathfrak n}l}$  as a function of $l$ and for the dominant ${\mathfrak n}=0$ terms. For polytropic stars with $P(\rho)=k\rho^{1+1/n}$ our results imply ${E}_{{\mathfrak n}l}\propto l^{-n}$ in the large-$l$ limit. For the tabulated FPS and AP4 EOS the energy scales as ${E}_{{\mathfrak n}l}\propto l^{-0.51}$ and ${E}_{{\mathfrak n}l}\propto l^{-0.54}$, respectively. For all models the NS radius $R\approx 12{\rm km}$, and the mass $M\approx 2 M_\odot$ and $M\approx 1.4 M_\odot$ for the polytropic models with $n=1/2$ and $n=1$, respectively, and $M\approx 1.5M_\odot$ and $M\approx 1.9M_\odot$ for the FPS and AP4 EOS, respectively.
}
\end{center}
\end{figure}

We compute ${E}_{{\mathfrak n}l}$ numerically up to $l=100$ for a variety of NS EOS, see Appendix~\ref{app:normalmodes} for details on the computation. The contribution of the overtones (${\mathfrak n}>0$) is negligible and in practice we consider only the fundamental ${\mathfrak n}=0$ modes. The multipolar contributions as functions of $l$ are shown in Fig.~\ref{fig:El} for typical configurations.

Our analysis reveals a remarkable property which, to the best of our knowledge, has never been discussed before. For very compact objects, 
the multipolar contribution ${E}_{{\mathfrak n}l}$ decays very slowly in the large-$l$ limit. For a polytropic EOS, $P(\rho)=k\rho^{1+1/n}$, we find the following behavior in the large-$l$ limit,
\begin{equation}
 {E}_{0l}\sim \frac{m_\PBH^2}{R}\frac{\gamma}{l^n} \qquad l\gg1\,,\label{El}
\end{equation}
where the constant $\gamma$ is of order unity and its precise value depends on the EOS and on the central density of the NS. Although Fig.~\ref{fig:El} shows only the cases $n=1$ and $n=1/2$, we verified Eq.~\eqref{El} for $0\leq n\leq 3$. This result is not an artifact of the polytropic-fluid approximation. Indeed, we have considered two representative tabulated EOS, namely FPS~\cite{Friedman:1981qw} and AP4~\cite{Akmal:1998cf}, obtaining a result analogous to Eq.~\eqref{El} but with $n\approx 0.51$ and $n\approx 0.54$, respectively. This is natural because NS equilibrium structures are roughly described by polytropes with $0.5<n<1$.

Equation~\eqref{El} implies that the total energy loss $\Delta E\sim m_\PBH^2/R$ for polytropes with $n>1$, which are simple approximations of Sun-like stars and white dwarfs. On the other hand, $\Delta E=\sum_{{\mf n}l}{E}_{{\mathfrak n}l}$ formally \emph{diverges} for polytropes with $n\leq1$ and for realistic NS models. A similar phenomenon occurs in the head-on collision of a point-particle with a black hole. In that case, the energy \emph{absorbed} at the horizon diverges~\cite{Davis:1972ud}. Our results show that a similar divergence also occurs if the central object is a NS. This divergence is enabled by the fact that a PBH does not suffer tidal disruption; instead it reaches the radius of the star and even travels within it. In fact, an analysis of the excitation coefficients~\eqref{cext} and \eqref{cint} reveals that the majority of the energy is released when the particle reaches the surface and when it passes through the outer layers of the NS, where large density gradients are present and where the corresponding large-$l$ eigenfunctions peak.
To compute the eigenfunctions in the low-density region precisely, tabulated EOS were also implemented using piecewise functions~\cite{Read:2008iy}. In Section~\ref{sec:crust} we present a simple, analytical toy model that sheds some light on this peculiar behavior.

Although during the inside motion the energy loss is roughly an order of magnitude larger than during the outside motion, the divergence of $\Delta E$ would also occur if only the outside motion is taken into account, provided the PBH can reach the surface without being destroyed.

For an incompressible fluid with $n=0$, Eq.~\eqref{El} predicts ${E}_{0l}\approx {\rm const}$ in the large-$l$ limit, precisely as in the black-hole case~\cite{Davis:1972ud}. In the next sections we solve the $n=0$ case analytically, confirming that the divergence of the total energy is not a numerical artifact. We show that the energy deposited during the external infall reads $E_{0l}\to\frac{3}{4\pi}{m_\PBH^2}/{R}\sim0.24\, {m_\PBH^2}/{R}$, in the large-$l$ limit. This number is in good agreement with the black-hole case, in which $E_{0l}\sim 0.5\,{m_\PBH^2}/{R_S}$, where $R_S$ is the Schwarzschild radius~\cite{Davis:1972ud}.

\subsection{Tidal heating and dynamical friction}
Before entering the details on how the divergence of the total energy loss can be regularized, let us discuss the connection between our results and those of Ref.~\cite{Capela:2013yf}, in which the capture of a PBH by a NS through accretion and dynamical friction has been studied\footnote{The discussion of this section is in contrast with the recent comment by Capela et al.~\cite{Capela:2014qea}, which contains critical remarks about a preprint version of this paper. 
}. 

Let us start by discussing the differences between tidal heating~\cite{1977ApJ...213..183P,1986ApJ...310..176L,1994ApJ...426..688R,Lai:1993di} (the process considered in this paper) and dynamical friction~\cite{Chandrasekhar:1943ys,1999ApJ...513..252O} (the process considered in Ref.~\cite{Capela:2013yf}). It has been suggested that these are two different ``approaches'' to compute essentially the same quantity, i.e. the energy loss by a point perturber in the encounter with a NS~\cite{Capela:2014qea}. 
Here we remark that dynamical friction and tidal heating are, in fact, two totally different phenomena. This claim is supported by, at least, three arguments:
\begin{enumerate}
 \item The energy loss due to tidal heating is not limited to the case in which the satellite travels through the star, but it also occurs when the satellite orbits \emph{outside} the star~\cite{1977ApJ...213..183P,1986ApJ...310..176L,1994ApJ...426..688R,Lai:1993di}. In such case, the local density near the satellite is vanishing and the force due to dynamical friction is zero (at least in its standard formulation~\cite{Chandrasekhar:1943ys,1999ApJ...513..252O,Capela:2013yf}, cf. e.g. Eqs.~(1) and (12) in Ref.~\cite{1999ApJ...513..252O}). Indeed, our analysis generalizes the study by Press \& Teukolsky~\cite{1977ApJ...213..183P} to the case in which the PBH can travel very close or within the star, but it does reduce to their case when the particle travels in the exterior. In this case, the energy loss due to tidal heating is nonvanishing, whereas that due to dynamical friction is zero.
 \item The energy deposited in oscillation modes is crucially associated with a \emph{confined} fluid. This is in contrast with the standard derivation of dynamical friction, in which the point particle is considered to travel through an infinite medium~\cite{Chandrasekhar:1943ys,1999ApJ...513..252O}. In particular, an infinite medium does not support spheroidal modes. In other words, if the radius $R$ of the
 star goes to infinity, the energy accumulated in the
modes is zero. This is in agreement with the fact that --~as we show later~-- the relevant modes can be roughly interpreted as \emph{surface} waves supported by the pressure gradients at the boundaries~\cite{1989nos..book.....U,Kokkotas:1999bd}. In the large-radius limit, the energy deposited in the modes becomes negligible with respect to the energy loss due to dynamical
friction. The latter is still present when the PBH travels through the star, regardless the existence of a surface.
 \item Similarly, in the limit in which the fluid is pressureless, no modes can be excited and the energy deposited into modes is again zero, whereas that associated to dynamical friction is not.
\end{enumerate}
These arguments demonstrate that the two mechanisms (tidal heating of the stellar modes and energy loss due to dynamical friction by an infinite, pressureless medium) \emph{are} very different in nature and, therefore, there is no reason to expect that the energy loss due to these effects would be equivalent or even comparable~\cite{Capela:2014qea}. The two effects exist independently from each other and, indeed, one would generically expect situations in which one of the two mechanisms is dominant\footnote{It would be interesting to incorporate the effects of a surface and of a nonvanishing fluid pressure into the standard calculation of the hydrodynamical drag~\cite{Chandrasekhar:1943ys,1999ApJ...513..252O}. In a sense, the computation presented here does precisely that, although using a completely different formalism. Likewise, it is reasonable to expect that the effects of standard dynamical friction are automatically included in our formalism. For instance, the large-${\mathfrak n}$ modes are strongly localized in the radial direction and the energy deposited into these modes might be interpreted as the counterpart of a local wake effect. Such analogy is beyond the scope of this paper but is certainly worth exploring further.}. Indeed, we stress that our results are not in contrast with those presented in Ref.~\cite{Capela:2013yf}, as the latter do not consider the energy deposited in the NS nonradial modes.

On the other hand, given the intrinsic different nature of the two effects, the causality argument presented in Ref.~\cite{Capela:2014qea} is unjustified. 
To further support this claim, we note that: (i) the same causality argument would exclude energy loss due to tidal heating in close parabolic and elliptical encounters, thus invalidating the well-accepted results by Press \& Teukolsky~\cite{1977ApJ...213..183P} (see also Refs.~\cite{1986ApJ...310..176L,1994ApJ...426..688R,Lai:1993di}); and (ii) from the point of view of tidal heating, the fact that perturbations in the fluid move at the speed of sound does not imply that the satellite cannot deposit energy into the modes during a single passage. During the first close encounter (either in the exterior or in the interior of the star) the satellite would immediately deposit some energy into the fluid. Only afterwards the energy propagates at the speed of sound and it is eventually dissipated on long time scales. Contrary to the case of drag forces, the energy loss due to tidal heating does not require back-reaction of the fluid onto the perturber (i.e. a wake effect) and the subsonic/supersonic nature of the satellite motion is irrelevant. This is particularly evident when the motion takes place in the exterior of the star, where the very notion of subsonic and supersonic motion is meaningless and, yet, tidal capture can occur~\cite{1977ApJ...213..183P,1986ApJ...310..176L,1994ApJ...426..688R,Lai:1993di}.

\section{Breakdown of the point-particle approximation and energy cutoffs}\label{sec:cutoff}
The divergence of the total energy loss as implied by Eq.~\eqref{El} suggests a breakdown of the point-particle approximation. 
Although this is an interesting result \emph{per se} and would deserve an independent analysis, here we are more interested in understanding how this formal divergence is regularized by physical processes occurring during the encounter.

\subsection{Finite-size effects}
In the black-hole case, it has been shown that a finite-size cutoff for the point particle regularizes such divergence~\cite{Davis:1972ud}. In practice, the multipolar sum~\eqref{seismic} should be truncated at
\begin{equation}
  l_{\rm max}\sim \frac{\pi}{2}\left(\frac{ R}{2m_\PBH}\right)\gg1\,, \label{cutoff}
\end{equation}
to account for finite-size effects on the scale of the PBH horizon~\cite{Davis:1972ud}. By applying the same cutoff in the NS case, the total energy is finite and reads
\begin{equation}
 \Delta E=  \frac{m_\PBH^2}{R} \frac{2\gamma}{(1-n)} l_{\rm max}^{1-n}\,, \label{energyloss}
\end{equation}
where we assumed $n<1$ and we stress that Eq.~\eqref{energyloss} applies also to realistic EOS which effectively correspond to $n\approx 0.5$. Note that this energy release is larger than the energy associated with accretion and dynamical friction~\cite{Capela:2013yf} by a factor $l_{\rm max}^{1-n}\gg1$. Even if the energy loss can be strongly amplified, $\Delta E$ is still much smaller than the PBH rest mass. This is in contrast with the black-hole case in which $\Delta E\sim m_\PBH$~\cite{Davis:1972ud}, signaling a possible breakdown of the perturbative approach. The case of realistic NSs with $n\sim 0.5$ is less problematic, because after regularization $\Delta E\ll m_\PBH$ and, individually, each contribution $E_{{\mathfrak n}l}$ is much smaller than the dominant quadrupole, cf. Fig.~\eqref{fig:El}. This suggests that the linear analysis is accurate and nonlinear corrections to the high-$l$ modes should be small.

The cutoff~\eqref{cutoff} is obtained by requiring the \emph{angular} resolution to be smaller than the PBH size, so that the point-particle approximation holds. A different cutoff is associated to the same requirement in the \emph{radial} direction, which can be expressed by imposing that the wavelength of the mode be much larger than the PBH size, i.e. $\omega_{{\mathfrak{n}}l}\ll\pi v_{\rm sound}/m_\PBH$, where $v_{\rm sound}$ is the speed of sound inside the NS and we assumed a linear dispersion relation for the sound waves. It is well-known that the frequency of the $\mathfrak{n}=0$, f-modes depends almost exclusively on the macroscopic properties of the star,
\begin{equation}
 \omega_{0l}^2\sim\frac{l(l-1)}{l+1/2}\left(\frac{M}{R^3}\right)\,, \label{fmode}
\end{equation}
in analogy with the surface-gravity waves of a fluid~\cite{1989nos..book.....U,Kokkotas:1999bd}. Therefore, in the large-$l$ limit, the condition $\omega_{0l}\ll \pi  v_{\rm sound}/m_\PBH$ implies
\begin{equation}
 l\ll l_{\rm max}^{\rm rad}=\pi^2 v_{\rm sound}^2 \frac{R}{M} \left(\frac{R}{m_\PBH}\right)^2\,.\label{cutoffradial}
\end{equation}
Since $v_{\rm sound}\gtrsim 10^{-3}$ even in the outer layers of the NS structure this cutoff is negligible relative to Eq.~\eqref{cutoff} in the region $m_\PBH/M\lesssim 10^{-8}$ under consideration\footnote{When the central object is a black hole, the frequency of the fundamental mode in the large-$l$ limit is linear in $l$, $\omega_{0l}\sim l$~\cite{Berti:2009kk}, and the mode propagates at the speed of light. In this case the radial cutoff is comparable to Eq.~\eqref{cutoff}, $l_{\rm max}^{\rm rad}\sim l_{\rm max}$.}.

Nonetheless, in the interior of a NS other scales (associated to a smaller cutoff than Eq.~\eqref{cutoff}) can be relevant. The latter are discussed in detail in the subsection below. Estimating the minimum cutoff for this problem is particularly important because, in the mass range $m_\PBH<10^{24}{\rm g}$ under consideration, the radius of the infalling PBH is of the order of the microns which corresponds to a very large finite-size cutoff, $l_{\rm max}>10^{10}$.

\subsection{The effects of a core-crust interface}\label{sec:crust}
It has been pointed out that a core-crust transition would introduce a cutoff in $l$, so that our Eq.~\eqref{El} would be valid only up to some cutoff value, $l_d\sim R/d$, where $d$ is the thickness of the crust~\cite{Capela:2014qea}. Before discussing this important point, we note here a logical inconsistency with this type of arguments. The putative correspondence claimed in~\cite{Capela:2014qea} between standard dynamical friction and our computation would imply that our analysis should recover the dynamical friction formula in \emph{any} situations, including idealized ones. However, the argument then continues by making a very specific assumption, namely the existence of a core-crust transition. In the limit in which the crust thickness is zero, $d\to0$, the analysis of Ref.~\cite{Capela:2014qea} is not in contrast with ours.

Now, let us consider a \emph{thought} experiment in which we study the energy loss due to a point particle that travels through a hypothetical sphere of gas with polytropic index $n=1/2$. According to the analysis of Ref.~\cite{Capela:2014qea}, our result~\eqref{El} should be valid in principle for any $l$, because in this idealized case the crust is absent. This remains true even if in this case the motion in the interior of the sphere is always supersonic with $v_p > v_{\rm sound}\sim0.4$.  Nonetheless, in this case the energy accumulated into spheroidal modes would be dramatically different from the energy dissipated due to dynamical friction. The argument presented in Ref.~\cite{Capela:2014qea} cannot resolve this conundrum.
Clearly, this apparent inconsistency can be resolved if the two effects, dynamical friction and tidal heating, are considered as distinct, as argued above.
Leaving this issue aside, we now simply focus on the energy loss due to tidal heating and discuss an important issued related to the analysis of Ref.~\cite{Capela:2014qea}. Namely, does the existence of a solid crust in the outer region close to the NS surface affect the total energy loss? 

In Ref.~\cite{Capela:2014qea} it was argued, using a Cowling approximation and an eikonal expansion, that the spheroidal f-modes which dominate the energy loss peak at $\sim R(1-1/l)$ and, therefore, when $l>l_d\sim R/d$ such modes have support only in the crust region and cannot depend on the EOS of the NS core.

The conclusion of Ref.~\cite{Capela:2014qea} is based on a \emph{local} analysis which neglects the boundary conditions that have to be imposed at the core-crust layer and can change dramatically the conclusions. This is not surprising, since the eigenvalue problem that defines the eigenfunctions and the normal modes is fully characterized only after suitable boundary conditions are imposed. The latter should not only be imposed at the radius and at the center, but \emph{also} at the interface layer, if any.

To investigate this issue more quantitatively, in Appendix~\ref{app:normalmodes} we present a toy model using a two-component star made of two incompressible fluids with different constant density, merged together at some transition layer $r=R_1<R$. At $r<R_1$ the density $\rho=\rho_c=$const, whereas at $r>R_1$ the density jumps to $\rho_1\ll\rho_c$.
To the best of our knowledge, this problem is not discussed in textbooks, although it can be remarkably solved in closed form for any $l$ and shows several interesting properties. We make the assumption of an incompressible fluid only for heuristic purposes (because the analytic resolution allows to control the large-$l$ regime more easily), but the qualitative result would be the same if realistic EOS are adopted. In a more realistic situation, the crust has to be described by a solid material with a certain shear tensor (cf. e.g. Refs.~\cite{1988ApJ...325..725M,Kruger:2014pva} and Sec.~\ref{sec:damping}). The Newtonian models presented here are exact, i.e. we do not adopt a Cowling approximation nor a large-$l$ expansion.

In Section~\eqref{sec:2fluid} we show that the eigenvalue problem reduces to a second-order algebraic equation for the quantity $\omega_l^2$, which schematically reads
\begin{equation}
 a_2 \omega_l^4+a_1\omega_l^2+a_0=0\,, \label{modes2}
\end{equation}
where the coefficients $a_i$ are complicate functions of the model parameters. In general, for each $l$ the system admits more than one single eigenmode. This is in contrast with the single incompressible fluid case whose eigenfrequencies are given by Eq.~\eqref{wKelvin} and are in one-to-one correspondence with $l$. In this two-component model, for each $l$ we find two eigenfunctions which behave very differently from each other.

 \begin{figure}[t]
 \begin{center}
\epsfig{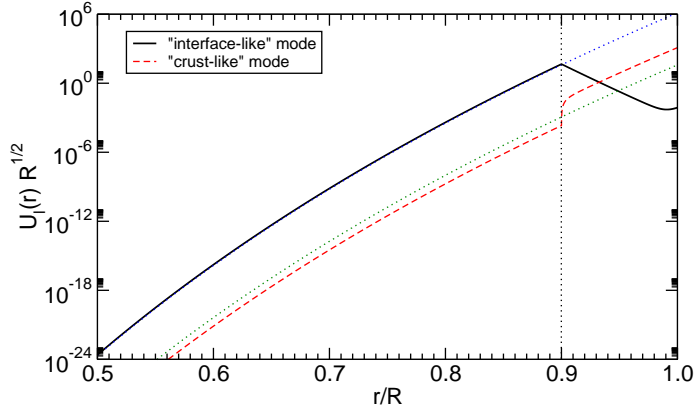}
\caption{\label{fig:eigenfunctions}
Radial displacement for a two-component incompressible fluid model with $\rho_1=10^{-3} \rho_c$, $R_1=0.9 R$, $R=5M$ and $l=100$. The eigenvalue problem admits two solutions with different properties. ``Interface-like'' modes peak at the interface layer, $r=R_1$ (marked with a vertical dotted line). In the large-$l$ limit, these modes have support only near the interface. On the other hand, ``crust-like'' modes grow monotonically and they have support mostly in the outer region. The dotted curves represent the radial displacement of a single-fluid model, Eq.~\eqref{UV_incompressible}, with the same total mass and radius $R_1$ (top curve) and $R$ (bottom curve), respectively.
}
\end{center}
\end{figure}

A representative example is shown in Fig.~\ref{fig:eigenfunctions}. In this example, we have fixed $\rho_1=10^{-3} \rho_c$, $R_1=0.9 R$, $R=5M$ and $l=100$. We find two distinct families of modes, which we label ``interface-like'' and ``crust-like'' modes in Fig.~\ref{fig:eigenfunctions}. As a consequence of the boundary conditions at the layer, the former family peaks at the interface, $r=R_1$, whereas the second family peaks in the outer region, precisely like the single-fluid eigenfunctions~\eqref{UV_incompressible}. Indeed, the dotted curves in Fig.~\ref{fig:eigenfunctions} represent the radial displacement~\eqref{UV_incompressible} for a single-fluid model, with the same total mass and radius $R_1$ and $R$, respectively. It is important to note that the interface-like modes match perfectly the single-fluid result only up to the interface $R_1$, whereas they behave very differently for $r>R_1$. 

The presence of more modes is natural, because the system can support surface modes at \emph{both} boundaries, $R_1$ and $R$, similarly to the well-studied interface modes\footnote{Another relevant paper in this context is Ref.~\cite{Tsang:2011ad} where --~using a formalism very similar to the one adopted here~-- NS crust shattering is explained in terms of energy deposited in core-crust interface modes due to a close NS-NS encounter.} in realistic NS models, which are related to the presence of a discontinuity layer~\cite{Kokkotas:1999bd}.

Most importantly, Fig.~\ref{fig:eigenfunctions} shows that the interface-like eigenfunctions peak at the interface layer. This property is true for any value of $l$ and it is a consequence of the boundary conditions imposed at the interface. Indeed, the radial displacement reads
\begin{equation}
 U_l(r)=\left\{ \begin{array}{l}
                 A r^{l-1} 			\qquad \hspace{3.2cm} r<R_1 \\
                 C_1 r^{-(l+2)}+ C_2 r^{l-1} \qquad R_1<r<R
                \end{array}\right.\,,
\end{equation}
where the constants $C_i$ and $A$ are defined in Section~\eqref{sec:2fluid}. For typical values of the integration constants, the radial displacement presents a local maximum at $r\sim R_1$ whose height grows in the large-$l$ limit.
This is in contrast with the analysis of Ref.~\cite{Capela:2014qea}, because the latter does not take into account the boundary conditions at the interface layer.

\begin{figure}[t]
 \begin{center}
\epsfig{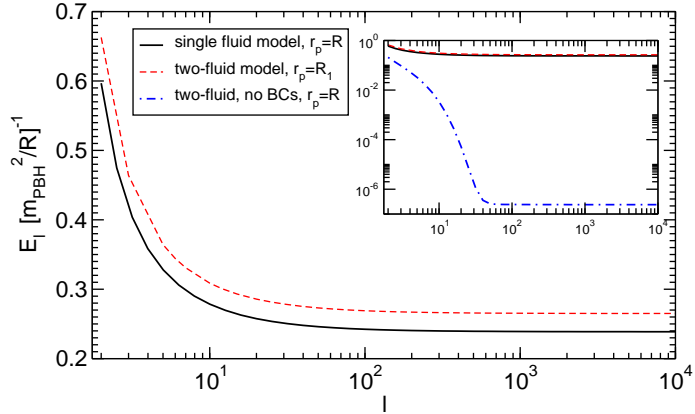}
\caption{\label{fig:El_2fluid}
Multipolar contributions ${E}_{l}$ to the total energy $\Delta E=\sum_{l}{E}_{l}$  as a function of $l$ for some models of incompressible fluids and with the same parameters as in Fig.~\ref{fig:eigenfunctions}. The solid black line represents the result of a single-fluid model with static point-like source located at $r_p=R$, whereas the red dashed line shows the analogous result for our two-fluid model with $r_p=R_1$. In both cases, the energy ${E}_{l}$ is constant in the large-$l$ limit. The inset shows the same result but including the energy loss due to an artificial single-fluid mode that peaks in the crust at large $l$; see text for details.
}
\end{center}
\end{figure}

Finally, let us discuss the energy loss due to a static point source, which is a simple toy model a  often used in seismology~\cite{seismology,Luo:2012pp} (cf. Section~\ref{sec:static} for details). Despite its simplicity, this model retains the key features of the phenomenon at hand. Using the eigenfunctions computed above, we can directly apply Eqs.~\eqref{cext2} and \eqref{cint2}. A representative result is shown in Fig.~\ref{fig:El_2fluid}.
In this figure, we show the energy ${E}_{l}$ accumulated in the $l$-th mode as a function of $l$ up to $l=10^4$ and for some models of incompressible fluids with the same parameters as in Fig.~\ref{fig:eigenfunctions}. The solid black line represents the result of a single-fluid model where the point source is located at the radius $r_p=R$, whereas the red dashed line shows the analogous result for our two-fluid model where the point source is located at the interface $r_p=R_1$. In both cases, the energy ${E}_{l}$ is constant in the large-$l$ limit, in agreement with our Eq.~\eqref{El}. Indeed, the result is very similar for both cases, the small difference being due to the fact that the two models have the same total mass, $M(R)$, but the interface-like modes of the two-fluid model are effectively equivalent to the single-fluid modes of a more compact star, because they peak at $R_1<R$ (cf. Fig.~\ref{fig:eigenfunctions}), and are almost insensitive to the outer region.

Note that, because we chose $R_1=0.9R$ for heuristic purposes, the thickness of the outer region is $d\sim0.1 R$ and, according to the analysis of Ref.~\cite{Capela:2014qea}, our Eq.~\eqref{El} should break at $l\gtrsim l_d\sim 10$. However, such expectation is based on the fact that high-$l$ modes peak at $r\sim R(1-1/l)$, whereas we have just shown that, when boundary conditions are properly taken into account, the relevant modes peak at $r\sim R_1$ for any $l$. Indeed, our results are consistent with Eq.~\eqref{El} even for $l=10^4$ (and in fact even for higher $l$), which is much higher than the cutoff value $l_d$ derived in Ref.~\cite{Capela:2014qea}.

In order to recover the results of Ref.~\cite{Capela:2014qea}, we can \emph{artificially} neglect the boundary conditions at the layer, for example using a background two-fluid model, but insist on using the single-fluid eigenfunctions given in Eq.~\eqref{UV_incompressible}. The result for the energy loss in this case is shown in the inset of Fig.~\ref{fig:El_2fluid}. For large values of $l$, the eigenfunctions have support only at $r\sim R(1-1/l)$, as argued in Ref.~\cite{Capela:2014qea}. Indeed, in this case the energy dramatically drops down when $l\gtrsim l_d\sim R/(R-R_1)\sim10$, because high-$l$ modes peak in the region of low density $\rho_1\ll \rho_c$. We stress, however, that such modes do not exist in the spectrum, because they do not satisfy the boundary conditions at the interface layer. A similar result holds for the crust-like modes shown in Fig.~\ref{fig:eigenfunctions}. The energy deposited in these modes is however negligible with respect to the dominant contribution that comes from the interface-like modes, so neglecting them does not affect the total energy loss.

To summarize, our analysis shows that the presence of a crust-core transition does not change qualitatively the total energy loss presented in Section~\ref{sec:tidal}. This is due to the fact that the relevant modes peak at the transition layer and are therefore more sensitive to the interior of the NS rather than the crust. This property was missed in Ref.~\cite{Capela:2014qea} because the latter analysis ignores the boundary conditions at the interface layer.

Finally, we remark that although we used a simple incompressible fluid model with a static point source, our results would qualitatively hold also in the case of more realistic EOS and for a moving source. In particular, the maximum energy loss would correspond to the PBH travelling through the region where the relevant modes peak, i.e. when $r_p(t)\sim R_1$. The existence of NS normal modes that peak at the core-crust interface is well known~\cite{1989nos..book.....U,Kokkotas:1999bd} (see also Ref.~\cite{Tsang:2011ad} for a recent application). The entire analysis of this section can be repeated for a moving source with qualitatively similar results, although in this case numerical estimates of the excitation coefficients are required.

\subsection{Damping mechanisms associated to elasticity and viscosity}\label{sec:damping}

At least two further mechanisms can potentially affect the energy deposited in high-$l$ modes, namely the presence of a solid crust and damping due to viscosity.

To include the effects of the former, the fluid describing the NS core has to be interfaced to a solid crust, whose properties are described by the shear modulus $\mu$ of the material, typically iron~\cite{1988ApJ...325..725M,Kruger:2014pva}. Besides affecting the boundary conditions at the core-crust interface, elasticity of the crust also changes the dynamical equations that describe the eigenfunctions in the outer layers. Since our approach neglects elasticity, our results are strictly valid only when the effects of a solid crust are small. This happens when the ratio $\mu/\rho$ is much smaller than the speed of sound~\cite{1988ApJ...325..725M,Kruger:2014pva}.

From standard assumptions on the physics of the NS crust~\cite{Chamel:2008ca}, we estimate that $\mu/\rho\sim v_{\rm sound}$ at density $\rho\sim10^5 {\rm g/cm}^3$, which corresponds to the
very-outer crust region roughly one millimeter beneath the surface. Using the same argument that brings to Eq.~\eqref{cutoffradial}, this condition implies the cutoff
\begin{equation}
 l\ll l_{\rm max}^{\rm elast}\sim 3\times 10^{10}\left(\frac{v_{\rm sound}}{10^{-3}}\right)^2\left(\frac{R}{5M}\right)\left(\frac{R}{12\,{\rm km}}\right)^2\,.\label{cutoffshear}
\end{equation}
Thus, the effects of the crust elasticity are negligible for modes up to $l\sim 10^{10}$. We note that Eq.~\eqref{cutoffshear} does not necessarily imply that modes with $l>l_{\rm max}^{\rm elast}$ are not excited during the infall, but only that the amplitude of such modes has to be computed including the crust elasticity in the eigenvalue problem~\cite{1988ApJ...325..725M,Kruger:2014pva}.

As for the effects of viscosity, the latter become relevant when their corresponding time scale is comparable to the PBH crossing time, $\tau_c\sim 4\times 10^{-5} v_p [R/(12\,{\rm km})]\,{\rm s}$, where the PBH escape velocity $v_p\sim 0.6$ near the NS. Simple expressions for the dissipative time scales as functions of $l$ and the NS parameters where derived in Ref.~\cite{1987ApJ...314..234C}:
\begin{eqnarray}
 \tau_\eta&=&\frac{1.3\times 10^{10}}{(l-1)(2l+1)} \left(\frac{10^{14}\,{\rm g/cm}^3}{\rho}\right)^{5/4} \left(\frac{T}{10^9\,{\rm K}}\right)^{2} \left(\frac{R}{12\,{\rm km}}\right)^{2} \,{\rm s}\,,\\
 \tau_\kappa&=& 5.5\times 10^6\tau_\eta \frac{(l-1)^2}{l^3} \left(\frac{\rho}{10^{14}\,{\rm g/cm}^3}\right)^{19/12} \left(\frac{T}{10^9\,{\rm K}}\right)^{-2} \left(\frac{R}{12\,{\rm km}}\right)^{2}\,,\\
 \tau_\zeta&>&61 \tau_\eta \frac{\eta}{\zeta}\,,
\end{eqnarray}
where $T$ is the NS temperature, $\tau_\eta$ $\tau_\kappa$ and $\tau_\zeta$ are the time scales for shear viscosity, thermal conductivity and bulk viscosity, respectively, whereas $\eta$, $\zeta$ and $\kappa$ are dissipation coefficients. 
The bulk viscosity coefficient $\zeta$ is typically comparable to $\eta$, so that its time scale is always longer than that associated to shear viscosity. Although in the interior of a NS the shear viscosity is typically the dominant dissipation mechanism~\cite{1987ApJ...314..234C}, the relative relevance of $\tau_\eta$ and $\tau_\kappa$ depends on the local density and temperature of the region where the modes peak and on the multipolar index $l$.

Imposing $\tau_c\ll \tau_\eta$ and $\tau_c\ll\tau_\kappa$ and solving for $l$, we obtain the cutoffs
\begin{eqnarray}
 l\ll l_{\rm max}^{\rm \eta}   &\approx& 10^7   \left(\frac{10^{14}\,{\rm g/cm}^3}{\rho}\right)^{5/8} \left(\frac{T}{10^9\,{\rm K}}\right) \left(\frac{R}{12\,{\rm km}}\right) \,,\label{cutoffeta}\\
 l\ll l_{\rm max}^{\rm \kappa} &\approx& 10^7  \left(\frac{\rho}{10^{14}\,{\rm g/cm}^3}\right)^{1/9} \left(\frac{R}{12\,{\rm km}}\right)^{4/3}\,.\label{cutoffkappa}
\end{eqnarray}
Note that $l_{\rm max}^{\rm \kappa}$ is independent from the NS temperature and decreases (although very mildly) at small densities, whereas $l_{\rm max}^{\rm \eta}$ is linear in $T$ and increases at small densities. It is easy to check that for typical values of the NS crust~\cite{Chamel:2008ca} --~and even for densities as low as $\rho\sim 10^5\,{\rm g/cm}^3$~-- the cutoffs derived above always exceed one million.

To summarize, finite-size effects of the source, the elasticity of the crust and viscosity introduce various cutoffs on the maximum harmonic index $l$. However, under rather conservative assumptions, we estimate that such effects give negligible contributions for modes with $l\lesssim 10^{7}$.
In light of these various effects and on the uncertainties on the NS structure, in the following we consider both the finite-size cutoff~\eqref{cutoff} and a much more conservative choice, $l_{\rm max}=10^6$, corresponding to a linear resolution at the NS surface of roughly $R/l_{\rm max}\sim {\rm cm}$, i.e. orders of magnitude larger than the PBH size and also more conservative than any other cutoff discussed in this section. We note that even in this very conservative case the energy loss~\eqref{energyloss} is roughly three orders of magnitude larger than in the case of dynamical friction.

\section{Tidal capture rate}\label{sec:capture}
During a collision with a NS, a PBH loses an amount of energy given by Eq.~\eqref{energyloss}, and might get gravitationally captured by the star. The time scale for this process and the associated capture rate were computed in Ref.~\cite{Capela:2013yf}, here we briefly summarize the analysis and refer to the original work for details (see also Ref.~\cite{Kouvaris:2007ay}).

Given a particle with energy $E_p$ in radial orbit with apastron $r_{\rm max}$, the half period of the orbit reads $\Delta T={\pi r_{\rm max}^{3/2}}/{\sqrt{M}}$. Due to the energy loss, $dE_p/dt\sim -\Delta E/\Delta T$, the equation of motion for the apastron reads
\begin{equation}
 \dot{r}_{\rm max}=-\frac{\Delta E}{\pi m_\PBH}\sqrt{\frac{r_{\rm max}}{M}}\,,\label{rdot}
\end{equation}
and $r_{\rm max}$ decreases at each encounter. Eventually, after several successive encounters, the PBH would be confined in the NS core in a time scale defined by $r_{\rm max}(t_{\rm loss})=0$. Solving Eq.~\eqref{rdot} one obtains
\begin{equation}
 t_{\rm loss}\sim 2 \pi\frac{ m_\PBH}{\Delta E} \sqrt{M r_{\rm max}^0}\,, \label{tloss}
\end{equation}
where $r_{\rm max}^0= r_{\rm max}(t=0)$ can be estimated by requiring that the PBH is gravitationally capture after the first encounter, i.e. by requiring that 
the initial energy is comparable to $\Delta E$. This translates in $r_{\rm max}^0\sim M m_\PBH/\Delta E\approx 10^7 \,{\rm km}$ for $m_\PBH\sim 10^{24}\,{\rm g}$, $l_{\rm max}\sim 10^6$ and typical values of $M$ and $R$. Based on Eq.~\eqref{energyloss}, the time scale~\eqref{tloss} depends on the NS configuration. For $n=1/2$, we obtain that only PBHs with 
\begin{equation}
 m_\PBH\gtrsim 10^{16}\left(\frac{10^6}{l_{\rm max}}\right)^{1/2}  \left(\frac{M}{1.4 M_\odot}\right)^{2/3}\left(\frac{R}{12\,{\rm km}}\right)\,{\rm g}\,, \label{thresholdmass}
\end{equation}
can be captured on a time shorter than a typical NS lifetime, $t_{\rm NS}\sim 10^{10}{\rm yr}$. Qualitatively similar results hold for other values of $n$ in Eq.~\eqref{energyloss}. The bound~\eqref{thresholdmass} shows that essentially any nonevaporating PBH is tidally captured on a time shorter than the NS lifetime, even using a very conservative cutoff, $l_{\rm max}\sim 10^6$.

The estimate above ignores possible interactions between the modes excited in a past passage and the motion of the PBH during the next passage. Resonances might occur when the orbital frequency $\Delta T^{-1}$ is comparable to the mode frequencies which, for the fundamental f-modes, is given in Eq.~\eqref{fmode} (see also Refs.~\cite{1989nos..book.....U,Kokkotas:1999bd}). This frequency is comparable to $\Delta T^{-1}$ when $l={\cal O}(1)$ and $r_{\rm max}\sim R$, i.e. for low multipoles and only when the PBH is already confined within the NS. Our analysis ignores possible resonances for which the energy loss would be amplified when $r_{\rm max}\lesssim R$, reducing the time scale for tidal capture. 
%

Let us now estimate the capture rate~\cite{Kouvaris:2007ay}. 
Assuming that each PBH --~with energy $E_p$ and angular momentum $J_p$~-- has a velocity relative to the NS which follows a Maxwellian distribution with dispersion $\sigma$, the capture rate can be obtained by considering the region of the $(E_p,J_p)$ parameter space corresponding to those orbits whose periastron is at most equal to the radius of the star and that lose enough energy to become gravitationally bound after the first encounter. The latter condition is simply $E_p<\Delta E$. The former condition can be obtained by analyzing the geodesic motion of a nonrelativistic particle in a Schwarzschild background. The final result for the capture rate is~\cite{Kouvaris:2007ay,Capela:2013yf}
\begin{equation}
 F=\sqrt{6\pi}\frac{\rho_\PBH}{m_\PBH}\frac{2M R}{\sigma(1-2M/R)}\left[1-e^{-\frac{3\Delta E}{m_\PBH\sigma^2}}\right]\,, \label{capturerate}
\end{equation}
where the mass density of PBHs can be defined in terms of the DM density through $\rho_\PBH=f_\PBH\rho_\DM$, with $f_\PBH$ being the DM fraction in PBHs. The factor $(1-2M/R)$ is the relativistic correction that takes into account gravitational focusing due to a compact central object~\cite{Kouvaris:2007ay}.
In the equation above, we have followed Ref.~\cite{Capela:2013yf} and did not approximate the exponential term as done in Ref.~\cite{Kouvaris:2007ay}. Hence, this capture rate is valid for any ratio $\Delta E/(m_\PBH\sigma^2)$, see discussion in Ref.~\cite{Capela:2013yf} for details.

After being captured, a PBH with a mass satisfying Eq.~\eqref{thresholdmass} would reach the center of the star in a time $t_{\rm loss}\ll t_{\rm NS}$ and would quickly disrupt the star by rapid accretion~\cite{Giddings:2008gr}. 
Therefore, the mere existence of old NSs in DM-rich galaxies implies that the survival probability, $\exp(-t_{\rm NS}F)$, is not small. This translates to an upper bound on the DM fraction in PBHs~\cite{Capela:2013yf}:
\begin{equation}
 f_\PBH < \frac{m_\PBH \sigma(1-2M/R)}{2\sqrt{6\pi}t_{\rm NS} MR \rho_\DM}\left[1-e^{-\frac{3\Delta E}{m_\PBH\sigma^2}}\right]^{-1}\,. \label{final}
\end{equation}

\section{Limits on DM fraction in PBHs}\label{sec:limits}
Given a detection of an old NS in a DM-rich environment, Eq.~\eqref{final} can be used to put a constraint on the DM fraction in PBHs. This constraint is inversely proportional to the DM density $\rho_\DM$. Because slow PBHs are captured more easily, the most stringent constraints can be derived from galaxies with a low velocity dispersion and a high DM density. 

\begin{figure}[t]
 \begin{center}
\epsfig{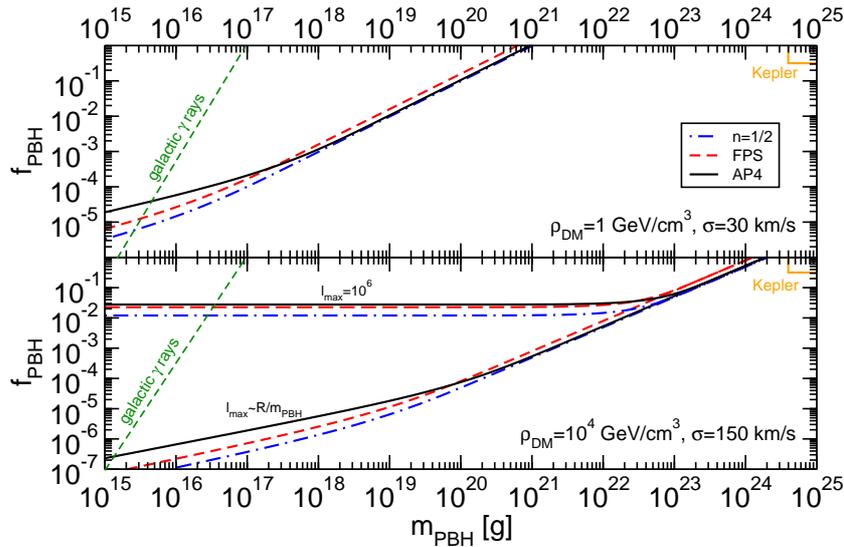}
\caption{
Constraints on the DM fraction in PBHs $f_\PBH$, derived from PBH capture by a NS in the PBH mass range $10^{15}{\rm g}\lesssim m_\PBH\lesssim 10^{25}{\rm g}$ currently unconstrained by observations. We consider a polytropic EOS with $n=1/2$ and the FPS and AP4 tabulated EOS with the same parameters as in Fig.~\ref{fig:El}. 
We consider typical values of the DM density $\rho_\DM$ and velocity dispersion $\sigma$ for the centers of the Large Magellanic Cloud~\cite{Alves:2000re} (top panel) and the Milky Way~\cite{DMcoresBin} (bottom panel). The orange rightmost top curve shows the existing limit from Kepler~\cite{Griest:2013aaa}, whereas the green leftmost line is the limit imposed by observations of the extragalactic photon background~\cite{Carr:2009jm}.
In the bottom panel, we show two different choices of the cutoff in $l$. The lower set of curves refers to $l_{\rm max}$ given by Eq.~\eqref{cutoff} as in the top panel, whereas the upper set of curves refers to $l_{\rm max}=10^6$. Even for this conservative choice, the theoretical bounds on $f_\PBH$ are still competitive at the level of few percent in the entire mass range.  
}\label{fig:constraints}
\end{center}
\end{figure}

In Fig.~\ref{fig:constraints} we show the exclusion plots for two representative examples: a region of moderate local DM density, $\rho_\DM\sim 1{\rm GeV~cm^{-3}}$, with low velocity dispersion $\sigma\sim 30{\rm km~s^{-1}}$, and a region of high DM density, $\rho_\DM\sim 10^4 {\rm GeV~cm^{-3}}$, but with high velocities, $\sigma\sim 150{\rm km~s^{-1}}$. The former choice is compatible with observations at about $1{\rm kpc}$ from the galactic center of the Large Magellanic Cloud~\cite{Alves:2000re}, whereas the latter corresponds to the center of the Milky Way in the conservative case of a mild DM spike~\cite{DMcoresBin}.
NSs are known to exist in the Large Magellanic Cloud~\cite{2013MNRAS.433..138R,2009ApJ...696..182N} as well as near the Milky Way center~\cite{Deneva:2009mx,2013Natur.501..391E}.

We stress that $\sigma$ in Eq.~\eqref{final} is the dispersion of the \emph{relative} velocity between the NS and the DM, so it depends on the intrinsic DM velocity dispersion and on the NS velocity distribution relative to the galactic center. The relative velocity between DM and the NS is expected to be typically smaller than the intrinsic DM velocity, unless the NS counter-rotates with respect to the DM halo.

To estimate how these constraints depend on the choice of $l_{\rm max}$ discussed in Sec.~\ref{sec:cutoff}, in the bottom panel of Fig.~\ref{fig:constraints} we also compare two different choices of the cutoff, namely Eq.~\eqref{cutoff} and a more conservative value $l_{\rm max}=10^6$. In this case the linear resolution at the NS surface is of the order of $1~{\rm cm}$, i.e. orders of magnitude larger than the PBH size and also more conservative than all cutoffs estimated in Sec.~\ref{sec:cutoff}. Even with this very conservative assumption, tidal capture can put competitive bounds at the level of few percent in the entire PBH mass range under consideration.

\section{Discussion \& Conclusions}\label{sec:conclusions}

Under normal circumstances, the energy deposited in the modes of a star during a close encounter with a PBH would be $\Delta E\sim m_\PBH^2/R$~\cite{1977ApJ...213..183P,1986ApJ...310..176L}, at most comparable to other effects such as energy loss due to accretion and dynamical friction~\cite{Capela:2013yf}. 
We have found that in a PBH-NS encounter the energy loss can be much higher and given by Eq.~\eqref{energyloss}. 
This has profound implications: nonevaporating PBHs can be trapped within the NS core in a time scale shorter than the NS lifetime and, if such capture occurs, the NS would quickly be destroyed through rapid accretion of its nuclear material onto the PBH~\cite{Capela:2013yf}. Thus, observations of NSs in DM-rich environments can be used to derive stringent constraints on the DM fraction in PBHs. 

Since tidal heating can be orders of magnitude larger than the energy dissipated through accretion and dynamical friction, the constraints that arise from tidal capture by a NS are more stringent and can be competitive also for high DM velocity dispersion or low DM density~\cite{Abramowicz:2008df}.

The absence of detection of microlensing events in the Kepler data --~together with previous bounds~\cite{Carr:2009jm}~-- sets the lower limit $m_\PBH\gtrsim 4\times 10^{24}{\rm g}$~\cite{Griest:2013aaa}, as shown by the yellow curve at the right-upper corner in Fig.~\ref{fig:constraints}. An analysis of Eq.~\eqref{final} shows that any NS detection in a region where the local DM density satisfies
\begin{equation}
 \rho_\DM\gtrsim 240 \left(\frac{\sigma}{{\rm km~s^{-1}}}\right){\rm GeV~cm^{-3}}\,, \label{limit}
\end{equation}
places new theoretical constraints on the DM fraction in nonevaporating PBHs with mass $m_\PBH\lesssim 4\times10^{24}{\rm g}$, i.e. in a region that is complementary to that excluded by Kepler. In deriving Eq.~\eqref{limit} we assumed $\sigma\ll 800\sqrt{l_{\rm max}/10^6}\,{\rm km\,s^{-1}}$ and a NS with $M\sim1.4 M_\odot$ and $R\sim 12{\rm km}$, but the prefactor would be of the same order for different choices of $M$ and $R$. As shown in the bottom panel of Fig.~\ref{fig:constraints}, the condition~\eqref{limit} is essentially independent from the cutoff $l_{\rm max}$.

As shown in Fig.~\ref{fig:constraints}, PBHs with $m_\PBH\lesssim 10^{17}{\rm g}$ are already excluded by observations of the extragalactic photon background~\cite{Carr:2009jm}. PBHs with $m_\PBH\gtrsim 10^{17}{\rm g}$ are larger than the size of a neutron by a factor of $\sim10^2$, implying that, when these objects interact with the NS, a fluid approximation of the NS structure is valid for matter with density as low as $\sim10^9 {\rm g~cm^{-3}}$, i.e. even in the outer layers near the NS surface. Smaller PBHs would instead interact with the single nucleons of the star in the outer layers, but such small PBHs are already ruled out because of their efficient Hawking evaporation~\cite{Carr:2009jm}.

Thus, our analysis excludes a range of more than seven orders of magnitude, $10^{17}{\rm g}\lesssim m_\PBH\lesssim 10^{24}{\rm g}$, which was previously viable. In combination with previous bounds, this suggests that PBHs of any mass cannot be the dominant constituent of the DM.

Our theoretical bounds depend rather strongly on the energy accumulated in the NS nonradial modes with large angular number $l$. A putative mechanism that makes the energy loss inefficient for modes with $l<10^6$ can in principle reduce the tidal heating and, in turn, make these constraints less stringent. We have discussed a variety of possible quenching mechanisms, and concluded that the latter should not affect the order of magnitude of our constraints. 

Furthermore, these bounds are only mildly dependent on the NS EOS and they are more sensitive to macroscopic properties such as the NS mass and radius. This is because they crucially depend on the energy deposited in the high-$l$ modes which, as we discussed, are mostly localized at the core-crust interface, where the behavior of matter is well understood and all realistic EOS are roughly equivalent.

In this work we have considered radial infall for simplicity. However, the enhancement of $\Delta E$ does not depend on this assumption and more generic motion (e.g. parabolic~\cite{1977ApJ...213..183P} or elliptical orbits~\cite{1994ApJ...426..688R,Lai:1993di}) would give rise to a similar effect as long as the periastron is close enough to the stellar radius or if the orbit crosses the NS. To illustrate this fact we adopted a simple toy model with a static source. In other words, the enhancement in the energy loss is not due to a ``focusing'' effect related to the fact that the PBH follows a radial orbit during the collision.

For closed orbits, tidal perturbations could stochastically excite the oscillation modes with random phases, so that the time-averaged amplitude of the displacement vector could be reduced due to destructive interference. However, because the mass ratio is extremely small, $m_\PBH/M\ll 10^{-8}$, the interaction between modes excited in the first passage and those in subsequent passages is suppressed, and each PBH passage encounters a nearly unperturbed NS. Interference between modes in multiple passages has been studied in Ref.~\cite{Li:2012gz}, showing that the mode energy grows linearly in time in stochastic repeated passages. Thus, even in case of interference the capture time for PBHs with $m_\PBH\gtrsim 10^{17}{\rm g}$ would remain shorter than the NS lifetime.

Our Newtonian approach provides a \emph{conservative} estimate, because it does not account for the energy loss due to gravitational-wave emission during the encounter, nor for the energy deposited in relativistic $w$-modes~\cite{Kokkotas:1999bd}. Taking these effects into account is an interesting extension of our work and would likely provide more stringent bounds than the ones we presented. In addition, a relativistic model can perhaps shed some light on the tidal-heating mechanism, whose explanation is interesting \emph{per se}, regardless its implications for DM searches. For instance, the fact that the energy loss due to tidal heating of a NS formally diverges might signal some profound properties of self-gravitating fluids governed by extreme EOS. Likewise, the similarity of this phenomenon with tidal heating of a black hole~\cite{Davis:1972ud} might have implications for the so-called "membrane paradigm"~\cite{Thorne:1986iy}.

The theoretical bounds derived here and in Ref.~\cite{Capela:2013yf} assume that the NS is nonspinning and, in particular, that the captured PBHs can accrete the nuclear material via spherical Bondi accretion. The effects of a spinning NS were recently investigated in Ref.~\cite{Kouvaris:2013kra}, finding that spin can have important effects for rapidly rotating and hot NSs. On the other hand, old NSs are cold and rotate rather slowly, so that the order of magnitude of our bounds and of those derived in Ref.~\cite{Capela:2013yf} should remain valid also in the moderately spinning case.

Finally, we note that there are currently no NS or pulsar detections in galaxies with very high DM densities and low velocity dispersion, such as the Milky Way's dwarf spheroidal satellites~\cite{2013pss5.book.1039W}. NS discoveries in these systems would place much tighter constraints than those presented here.

\acknowledgments
We thank Vitor Cardoso for valuable comments and Leonardo Gualtieri, Luis Lehner, Thomas Maccarone, Jocelyn Read, Jan Steinhoff and Jeroen Tromp for useful discussions. 
We also acknowledge interesting discussions with Andy Gould, Nicholas Stone, Fabio Capela, Maxim Pshirkov and Peter Tinyakov, which led to the expanded version of this manuscript.
P.P. was supported by the European Community through
the Intra-European Marie Curie contract aStronGR-2011-298297 and by FCT-Portugal through the projects IF/00293/2013 and CERN/FP/123593/2011. A.L. was supported in part by NSF grant AST-1312034.

\appendix

\section{Energy loss into spheroidal normal modes}\label{app:normalmodes}
\subsection{Eigenfunctions of a spherically-symmetric perfect fluid star}
The computation presented in Section~\ref{sec:tidal} is based on a normal-mode decomposition and requires to solve an eigenvalue problem for the normal modes of a spherically-symmetric perfect fluid NS. The latter is conveniently written in terms of five functions $y_i$ $(i=1,2,3,5,6)$, which satisfy the first-order system (cf. e.g.~\cite{1989nos..book.....U,GRL:GRL7294}):
\begin{eqnarray}
 y_1'&=& -\frac{2 y_1}{r}+\frac{y_2}{\kappa(r)}+\frac{l(l+1)}{r}y_3 \,,\\
 y_2'&=& -\left(\omega ^2\rho+\frac{4 \rho g}{r}\right)y_1+\frac{l(l+1)\rho g}{r}y_3-\rho y_6\,,\\
 y_3&=& \frac{1}{r \omega ^2}\left(g y_1-\frac{y_2}{\rho}-y_5\right)\,,\\
 y_5'&=& 4 \pi   \rho y_1+y_6\,,\\
 y_6'&=& -\frac{4 \pi   \rho 	l(l+1)}{r}y_3+\frac{l(l+1)}{r^2}y_5-\frac{2y_6}{r}\,,
\end{eqnarray}
where $g(r)=M(r)/r^2$ is the gravitational acceleration, $M'(r)=4\pi r^2 \rho$ and $\kappa=\rho dP/d\rho$ is the incompressibility.
In the notation of Ref.~\cite{GRL:GRL7294}, $y_1\equiv U_{{\mathfrak n}l}(r)$ and $y_3\equiv V_{{\mathfrak n}l}/\sqrt{l(l+1)}$, where $U_{{\mathfrak n}l}(r)$ and $V_{{\mathfrak n}l}(r)$ are the eigenfunctions used in the main text.
Note that $y_3$ is algebraically related to the other functions, so one is effectively left with a first-order system of four linear ordinary differential equations. 

The system above has to be solved by requiring regularity at $r=0$ and the boundary conditions
\begin{equation}
 y_2=0\,,\qquad y_6+\frac{l+1}{R}y_5=0 \,, \label{BCsR}
\end{equation}
at $r=R$. In addition, if the background fluid is made by more components, continuity boundary conditions have to be imposed at the junctions. In more realistic situations, the fluid has to be matched with a solid crust, described by a more general set of equations that include the shear modulus of the material. In such case, more complicated boundary conditions have to be imposed at the core-crust interface layer~\cite{1988ApJ...325..725M,Kruger:2014pva}.

After suitable boundary conditions are imposed, only a discrete set of frequencies $\omega_{{\mathfrak n}l}$ satisfy the eigenvalue problem. These are the normal modes of the star~\cite{1989nos..book.....U,Kokkotas:1999bd}. For a given harmonic index $l$, there is usually a countably infinite set of modes defined by the overtone number ${\mathfrak n}$ and associated to the number of nodes of $y_2(r)$. 
Finally, the eigenfunctions are canonically normalized such that
\begin{equation}
 \int_0^R dr\,r^2 \rho (U_{{\mathfrak n}l}^2+V_{{\mathfrak n}l}^2)=1\,, \label{orthonormality}
\end{equation}
and, in natural units, $U_{{\mathfrak n}l}$ and $V_{{\mathfrak n}l}$ have dimensions of ${\rm length}^{-1/2}$. Once the eigenfunctions are computed, Eq.~\eqref{seismic} together with Eqs.~\eqref{cext} and \eqref{cint} can be used to evaluate the energy deposited in the nonradial modes for a given source.

\subsection{Toy model: single-fluid incompressible sphere}\label{sec:incompressible}
As a toy model, let us consider a homogeneous ($\rho=\rho_c={\rm const}$), incompressible ($\kappa\to\infty$) sphere.
Remarkably this problem can be solved analytically, as shown by Lord Kelvin back in 1863. This single-fluid model is also propaedeutic for the computation presented in Sec.~\ref{sec:2fluid}. In this case the relevant normalized eigenfunctions are~\cite{seismology}
\begin{equation}
 U_{l}\equiv \frac{V_l}{\sqrt{1+1/l}}= \sqrt{\frac{4\pi l}{3M}}\left(\frac{r}{R}\right)^{l-1}\,, \label{UV_incompressible}
\end{equation}
and the eigenfrequencies read
\begin{equation}
 \omega_l^2=\frac{8\pi}{3}\frac{l(l-1)\rho_c}{2l+1}\,. \label{wKelvin}
\end{equation}
Note that, for each $l>0$, there is only one fundamental mode, ${\mathfrak n}=0$, and we shall omit the index ${\mathfrak n}$ for ease of notation.

With the eigenfunctions at hand, it is now straightforward to compute the energy accumulated in the modes due to an external source. 
For simplicity, let us consider the case of radial infall of a point particle outside the star up to the stellar radius $R$.
Using Eq.~\eqref{cext}, we obtain
\begin{equation}
 c_l^{\rm ext}=\frac{m_\PBH (l+1) \sqrt{M l} }{\sqrt{3 \pi } R^2}E_{N}(z)\,, \label{clincompressible}
\end{equation}
where $E_N(z)$ is the exponential integral function and we have defined $z=-\frac{2 i \sqrt{(l-1) l}}{3 \sqrt{2 l+1}}$ and $N=\frac{1}{3} (5+2 l)$. Using the equation above and Eq.~\eqref{seismic}, it is straightforward to show that
\begin{equation}
 {E}_{l}\sim \frac{3}{4\pi}\frac{m_\PBH^2}{R}\,, \quad {\rm for} \,\, l\gg1\,. \label{Elincompr}
\end{equation}
Therefore, the energy deposited in spheroidal modes is constant in the large-$l$ limit and the corresponding energy~\eqref{seismic} is divergent, as reported in the main text. Note that this divergence occurs for \emph{any} compactness. A similar result holds also for the internal motion. It is suggestive to note that the same type of divergence occurs in the collision of a point-particle with a black hole. In that case, the energy \emph{absorbed} at the horizon diverges because ${E}_l$ is constant in the large-$l$ limit~\cite{Davis:1972ud}. In fact, Eq.~\eqref{Elincompr} is also in quantitatively good agreement with the black-hole case. This peculiar result is due to the unrealistic assumption of incompressibility. As we shown in the main text, more realistic EOS --~which require numerical integration of the linearized equations~-- make such divergence much milder.
\subsection{Static source}\label{sec:static}
In order to simplify the problem even further, we can study a \emph{static} point-particle source located at $r=r_p$ and activated only at $t>0$. This is a common model often used in seismology~\cite{seismology,Luo:2012pp}. In this case, Eq.~\eqref{fsource} can be written as
\begin{equation}
 \boldsymbol{f}(\boldsymbol{x},t)\equiv -\Theta(t)\rho(\boldsymbol{x})\boldsymbol{\nabla}\Phi=  m_\PBH\Theta(t)\rho(\boldsymbol{x})\boldsymbol{\nabla}\frac{1}{|\boldsymbol{x}-\boldsymbol{x}_p|}\,,   \nn 
\end{equation}
where $\Theta(t)$ is the Heaviside function. The advantage of using a static source is that the excitation coefficients simplify considerably.
If $r_p\geq R$, we obtain
\begin{eqnarray}
 c_{{\mathfrak n}l}^{\rm ext}&=& \frac{m_\PBH  l}{r_p^{l+1}}\int_0^R dr\rho(r) r^{l+1}\left(U_{{\mathfrak n}l}+\frac{\sqrt{l(l+1)}}{l} V_{{\mathfrak n}l}\right)\,.\label{cext2}
\end{eqnarray}

In the case of an incompressible fluid with constant density, the energy accumulated in the $l$-th mode reads
\begin{equation}
 {E}_l^{\rm ext}=\left[\frac{3}{8\pi }\left(\frac{2 l+1}{l-1}\right)\left(\frac{R}{r_p}\right)^{2l+2}\right] \frac{m_\PBH^2}{R}\,,
\end{equation}
and, as $r_p\to R$, the energy becomes constant in the large-$l$ limit, similarly to the infall case previously discussed. Note also that the equation above is consistent with the result~\eqref{PT} derived in Ref.~\cite{1977ApJ...213..183P} where $r_p$ plays the role of the periastron distance. Indeed, such result confirms that the divergence in the total energy $\Delta E$ is not related to the radial infall, but it would also occur for more generic orbits, as discussed in the main text.

Likewise, if $r_p<R$, we obtain
\begin{eqnarray}
 c_{{\mathfrak n}l}^{\rm int}&&=  \frac{m_\PBH  l}{r_p^{l+1}}\int_0^{r_p} dr\rho(r) r^{l+1}\left(U_{{\mathfrak n}l}+\frac{\sqrt{l(l+1)}}{l} V_{{\mathfrak n}l}\right)\nn\\
 &&-m_\PBH (l+1)  r_p^{l} \int_{r_p}^Rdr\frac{\rho(r)}{r^l}\left(U_{{\mathfrak n}l}-\frac{\sqrt{l(l+1)}}{l+1} V_{{\mathfrak n}l}\right)\,. \label{cint2}
\end{eqnarray}
For an incompressible fluid, Eq.~\eqref{UV_incompressible} implies that the second line of the equation above vanishes identically. This is generically not true for more realistic backgrounds. In the case of an incompressible fluid, the energy accumulated in the $l$-th mode reads
\begin{equation}
 {E}_l^{\rm int}=\left[\frac{3}{8\pi }\left(\frac{2 l+1}{l-1}\right)\left(\frac{r_p}{R}\right)^{2l}\right] \frac{m_\PBH^2}{R}\,.
\end{equation}
Although this is an extremely simplified model, it contains all the relevant features discussed in the main text, namely when the source is localized near the NS radius (where the particle experiences large density gradients), the energy accumulated in the high-$l$ modes is largely amplified.

\subsection{Two-component incompressible fluid} \label{sec:2fluid}
Let us now present a model to quantify the relevance of interface layers for the total energy loss. Ideally, one should consider a model in which a perfect fluid with generic EOS $P(\rho)$ interfaces a solid material at some core-crust layer~\cite{1988ApJ...325..725M,Kruger:2014pva}. Here, in order to show the qualitative behavior of the eigenfunctions, we consider a much simpler model where two incompressible fluids are glued together at $r=R_1\lesssim R$. At $r<R_1$ the density $\rho=\rho_c=$const, whereas at $r>R_1$ the density jumps to $\rho_1\ll\rho_c$. Accordingly, the mass function reads
\begin{equation}
 M(r)=\left\{\begin{array}{l}
         ×\frac{4\pi}{3}r^3\rho_c \qquad             \hspace{2.7cm}  r<R_1 \\
          \frac{4\pi}{3}R_1^3\rho_c+\frac{4\pi}{3}r^3\rho_1 \qquad  R_1<r<R
        \end{array}\right. \,.
\end{equation}
Although such model is admittedly simplistic, it captures the large density jump that occurs at the core-crust interface~\cite{Chamel:2008ca}.

In the interior, $r<R_1$, the eigenfunctions that are regular at the center read
\begin{eqnarray}
 y_1&=& A r^{l-1}\,,   \nn  \\
 y_2&=& \rho_c\frac{r^l  \left(4\pi A l  \rho_c -3 B l-3 A \omega_l ^2\right)}{3 l}\,,   \nn  \\
 y_3&=& \frac{A}{l} r^{l-1}\,,   \nn  \\
 y_5&=& B r^l\,,    \nn \\
 y_6&=& r^{l-1} (B l-4 \pi  A \rho_c)\,,   \nn    
\end{eqnarray}
where $A$ and $B$ are integration constants. In the second region, $R_1<r<R$, the solutions are
\begin{eqnarray}
 y_1&=& C_1 r^{-2-l} + C_2 r^{l-1}  \,,    \nn \\
 y_2&=& \frac{1}{r^{l+1}}\left[\frac{4 \pi}{3}  \frac{R_1^3}{r^3} \rho_c \rho_1 \left(C_1+r^{2l+1} C_2\right)+r^{2l+1} C_3+C_4\right]\,,    \nn \\
 y_3&=& -\frac{l r^{-2-l} C_1-r^{l-1} C_2-l r^{l-1} C_2}{l+l^2}\,,   \nn  \\
 y_5&=& \frac{r^{-1-l}}{l (l+1) \rho_1} \left[l \rho_1 \left(\frac{4\pi}{3} (l+1)   \rho_1+ \omega_l ^2\right) C_1 +(l+1) r^{2l+1} \right.\nn\\
 &&\left.\times \left(\rho_1 \left(\frac{4\pi}{3} l  \rho_1- \omega_l ^2\right) C_2- l C_3\right)- l (l+1) C_4\right]\,,    \nn \\
 y_6&=& \frac{r^{-2-l}}{\rho_1} \left[\frac{4 \pi}{3}  \rho_1^2 \left((l-3) r^{2l+1} C_2-(4+l) C_1\right)\right.\nn \\
 &&\left.-\rho_1 \omega_l ^2 C_1-r^{2l+1} \left(\rho_1 \omega_l ^2 C_2+l C_3\right)+C_4+l C_4\right]\,,  \nn    
\end{eqnarray}
where $C_i$ are integration constants. Together with $A$ and $B$ we have 6 integration constants in total. Two of them can be fixed by imposing the boundary conditions~\eqref{BCsR} at the radius $r=R$. In addition, continuity of the functions $y_i$ at the interface $r=R_1$ imposes four additional conditions. These can be used to fix 3 constants plus the eigenfrequency $\omega_l$. Finally, the remaining integration constant is fixed by imposing the orthogonality condition~\eqref{orthonormality}. 
All these conditions are algebraic and, although cumbersome, they can be solved analytically and reduce (schematically) to Eq.~\eqref{modes2} in the main text.

\bibliographystyle{JHEP}
\bibliography{PBHs}
\end{document}